# Influence of the Cold War upon Influenza Pandemic of 1957-1958

Teruhiko Yoneyama
Multidisciplinary Science
Rensselaer Polytechnic Institute
Troy, New York, Unites States
yoneyt@rpi.edu

Mukkai S. Krishnamoothy
Computer Science
Rensselaer Polytechnic Institute
Troy, New York, United States
moorthy@cs.rpi.edu

*Abstract*—**Influenza Pandemic of 1957-1958, also called Asian Flu Pandemic, was one of the most widespread pandemics in history. In this paper, we model the pandemic, considering the effect of the Cold War. There were some restrictions between Western and Eastern nations due to the Cold War during the pandemic. We expect that such restrictions influenced the spread of the pandemic. We propose a hybrid model to determine how the pandemic spread through the world. The model combines the SEIR-based model for local areas and the network model for global connection between countries. First, we reproduce the situation in 19 countries. Then, we run another experiment to find the influence of the war in the spread of the pandemic; simulation considering international relationships in different years. The simulation results show that the impact of the pandemic in each country was much influenced by international relationships. This study indicates that if there was less effect of the Cold War, Western nations would have larger number of death cases, Eastern nations would have smaller number of death cases, and the world impact would be increased somewhat.**

I. INTRODUCTION

An influenza pandemic occurred during 1957-1958. This pandemic is called Asian Flu Pandemic and it is estimated that about 2 million people died during this pandemic [1][2][3][4]. This was one of the most widespread pandemics [5]. It is thought about 25-30% of the world's population experienced clinical disease and the mortality rate was estimated at approximately 1 in 4,000 [6]. The pandemic is believed to have originated from the China [7][8]. Table 1 shows the number of cases of the pandemic.

Table 1: Number of Cases of 1957-1958 Influenza by Country

|  | Deaths | Infected | Note | Source |
|---|---|---|---|---|
| US | 69,800 |  | Another Estimated Death Case is 80,000 [4][8][10] | [2][11][12] |
| Japan | 7,288 | 675,000 | Another Estimated Death Case is 5,700 [13] | [13] |
| UK | 6,716 |  |  | [1] |
| Singapore |  | >227,000 | By June 1957. Calculated from the description that 18% of population (1,261,677 in 1956 [5]). | [5] |
| Hong Kong |  | >488,000 | In early of pandemic. Calculated from the description that 15-20% of the population (2,440,000 in 1956 [5]). | [5] |
| Taiwan |  | 110,000 |  | [5] |
| Malaya |  | >80,000 | By the middle of May 1957 | [5] |
| India |  | >1,500,000 | By the middle of July 1957 | [5] |
| Philippines | 2,500 | >1,000,000 | By the end of June 1957 | [5] |
| Chile |  | 200,000 |  | [5] |
| World | ≈2 million | ≈820 million | Calculated from the description that a third of the world population (2,728,762,000 [5]) was affected [14]. | [4][14][15] |

Figure 1 shows the infection route of pandemic of 1957-1958. The pandemic started at China, and spread to Eastern European countries through Russia. Also, it spread to Western European countries as another infection route. The United States was infected through Japan. The infection was transmitted mainly through sea-lanes [8][9], and the pandemic had spanned the globe within approximately 6 months [8].

We hypothesize that the spread of the pandemic is based on the traffic pattern. Therefore, in this paper, we simulate 19 countries considering the international traffic in real data. This pandemic occurred during the Cold War. There were several restrictions between Western and Eastern nations. We expect such restrictions influenced the spread of pandemic. Thus we simulate, considering different international relationships in another year. By comparing this simulation result with the original simulation result, we can see how the Cold War influenced the pandemic.

To model the pandemic, we propose a hybrid model which considers both local infection and global infection. For the local infection, we use the SEIR model considering the each country's condition such as domestic population and population density. For the global infection, we use network based model considering the international traffic between countries.

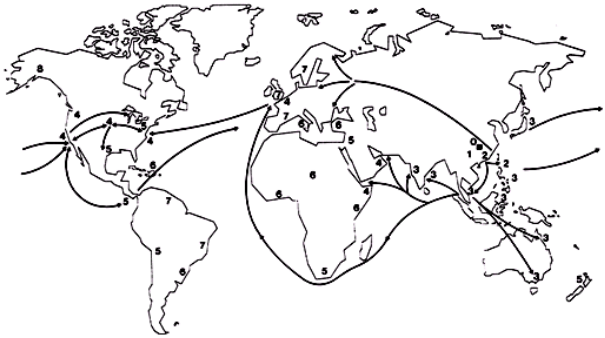

Figure 1: Infection Route of Influenza Pandemic of 1957-1958 [7][8]

## II. RELATED RESEARCH

Simulating the spreading of infectious disease has been studied in the past. We discuss the differences between this work and other related research. First, a lot of research about simulating disease spread focuses on a prevention/mitigation strategy by comparing the base simulation and an alternative simulation which considers their proposed strategy (e.g. [16][17][18][19][20][21]). In addition, most of existing research simulates with a generated situation which models the real world (e.g. [16][17][20][21][22][23] [24]). On the other hand, we focus on the reproduction of the real pandemic using real situation. We model the pandemic, compare the results with real data, and explore the key factors which influenced the spread. Although these critical-factors could provide hints that would help contain the spread of the disease, this paper does not directly propose a prevention strategy.

Second, much research considers the spread of infectious disease from either the local or global point of view (e.g. [18][20][21][23][25]). In addition, much research simulate using one of the equation based (e.g. SIR or SEIR differential equation model), agent based, or network based model (e.g. [23][26][27]). On the other hand, we simulate the pandemic from the global point of view considering local infection in each country. Also, we use a hybrid model which considers both the SEIR based model and network based model using the concept of agent based model.

Third, simulation parameters determine the path of spread. Some research values the basic reproduction number $R_0$ as an influential parameter (e.g. [18][28]). In our work, we don't determine $R_0$. In our simulation, we first consider setting the parameters so that the result corresponds with the actual situation in some countries in terms of the number of cases. Then we simulate further experiments using same set of parameters. This is based on the assumption that $R_0$ varies according to country.

## III. MODELING

Previous attempts to model spreading infectious diseases tended to fall into one of two categories. Equation-based models like the SEIR model is suitable for a large-scale spreading of diseases. These models use just a few parameters to reproduce the spreading phenomenon. However it is difficult to reflect detailed situation in countries which have different local infection conditions. Network or agent-based simulation models can theoretically reflect the detail of individual conditions. However, modeling large-scale global diseases is difficult as too many parameters are needed for simulation. Thus we propose a hybrid model. We make a simple model using a small number of parameters and make it capable of simulating a general pandemic.

We simulate using several countries. When we think of an infection in a country, there are three possibilities for new infection; (1) infection from foreign travelers, (2) infection from returning travelers, and (3) infection from local residents. Figure 2 illustrates this concept. We denote the infection-types (1) and (2) as the global infection and the infection-type (3) as the local infection.

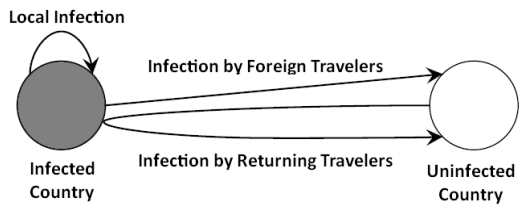

Figure 2: Three Patterns of Infection in a Country

We use the concept of SEIR model which considers four types of agents in each country; Susceptible, Exposed, Infectious, and Removed. Susceptible agents are infected by Infectious agents and become Exposed agents. Exposed agents are in an incubation period. After that period, Exposed agents become Infectious agents. Infectious agents infect Susceptible agents. Infectious agents become Removed agents after the infectious period. Removed agents are never infected again because they are now immune. Figure 3 illustrates this concept of SEIR model.

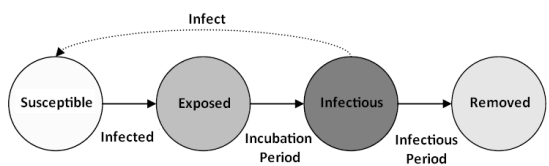

Figure 3: Concept of SEIR

At the beginning of the simulation, the number of Susceptible agents in each country is equal to the population of each country. Then we put an Infectious agent in the origin of the pandemic (i.e. China). The local infection spreads in the origin and the global infection also spreads from the origin to other countries through global traffic. When a country has at least one Infectious agent, that country has the potential for local infection. Figure 4 shows this concept.

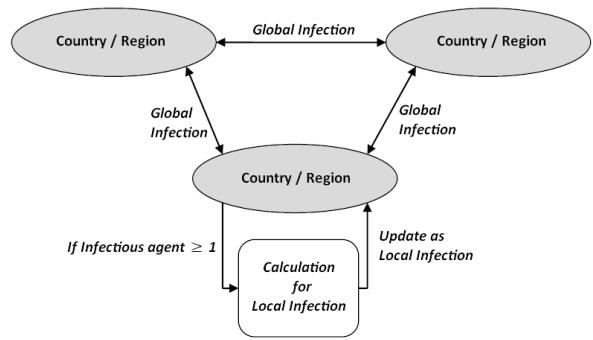

Figure 4: Concept of Simulation Task at One Cycle

The global infection is caused by traffic from infected country. Thus we refer to the number of inbound and outbound traffic. The number of new Exposed agents by the global infection in country $i$ at time $t$, $NEG_i(t)$, is calculated by the expression;

$$NEG_i(t) = I_j(t) \cdot T_{ij} \cdot P_G^*(t) \qquad (1)$$

where $I_j(t)$ is the number of Infectious agents of country $j$ at time $t$. $T_{ij}$ is the total amount of both traffic from country $i$ to $j$ and from $j$ to $i$. $P_G^*(t)$ is the global infection probability at time $t$ and is calculated by the expression;

$$P_G^*(t) = P_G - (D_G \cdot t) \qquad (2)$$

where $P_G$ is the basic global infection probability between countries. $D_G$ is a "deductor" for the global infection. $t$ is time (simulation cycle). $P_G$ and $D_G$ are constants and are uniformly used for every country. Thus the global infection probability $P_G^*(t)$, decreases along the simulation cycle. We assume that, in the real world, the global infection occurs with high probability in early pandemic due to the lack of awareness of the disease. As the disease spreads, people take preventive measures against the infection and the pandemic decreases. We apply this concept in the simulation. The number of Exposed agents in country $i$ at time $t$, $E_i(t)$, is updated by adding $NEG_i(t)$ to $E_i(t)$ at each simulation cycle.

We assume that the local infection probability depends on the population density of a country. Thus if the country is dense, people are more likely to be infected. The basic local infection probability of country $i$, $P_{Li}$ is given by the expression;

$$P_{Li} = Density_i \cdot C_1 + C_2 \quad (3)$$

where $Density_i$ is population density of country $i$, obtained by real data. Thus $Density_i$ differs in country. $C_1$ and $C_2$ are constants and are used for simulation in every country.

We assume that the number of new Exposed cases of a country by the local infection depends on the number of Susceptible agents and the number of Infectious agents at that time. Thus the number of new Exposed agents by the local infection in country $i$ at time $t$, $NEL_i(t)$, is calculated by the expression;

$$NEL_i(t) = S_i(t) \cdot I_i(t) \cdot P_{Li}^*(t) \quad (4)$$

where $S_i(t)$ us the number of Susceptible agents of country $i$ at time $t$. $I_i(t)$ is the number of Infectious agents of country $i$ at time $t$. $P_{Li}^*(t)$ is the local infection probability at time $t$ and is calculated by the expression;

$$P_{Li}^*(t) = P_{Li} - (D_L \cdot t) \quad (5)$$

where $P_{Li}$ is the basic local infection probability of country $i$ which is obtained by equation (3). $D_L$ is a "deductor" for the local infection and is a constant which is used for every country. $t$ is time (simulation cycle). Similar to the global infection, the local infection probability $P_{Li}^*(t)$ decreases as the simulation cycle increases. This reflects people's awareness. The number of Exposed agents in country $i$ at time $t$, $E_i(t)$, is updated by adding $NEL_i(t)$ to $E_i(t)$ at each simulation cycle.

Table 2 summarizes parameters in the simulation. We have eight controllable parameters which are denoted as constants in Table 2. These parameters are used for every country uniformly. Other parameters are derived from real data and depend on country.

Table 2: Parameters in Simulation

| Parameter | Description | Attribution | |
|---|---|---|---|
| | | (a)Global or (b)Local | (1) Constant or (2) Depend on Country |
| $P_G$ | Global Infection Probability | (G) | (1) |
| $P_{Li}$ | Local Infection Probability of County $i$ | (L) | (2) |
| $D_G$ | Deductor for Global Infection Probability | (G) | (1) |
| $D_L$ | Deductor for Local Infection Probability | (L) | (1) |
| $C_1$ | Constant for Local Infection Probability | (L) | (1) |
| $C_2$ | Constant for Local Infection Probability | (L) | (1) |
| Incubation_Period | Incubation Period | (G) and (L) | (1) |
| Infectious_Period | Infectious Period | (G) and (L) | (1) |
| Run_Cycle | Run Cycle of Simulation | (G) and (L) | (1) |
| $Density_i$ | Actual Population Density of Country $i$ | (L) | (2) |
| $Population_i$ | Actual Population of Country $i$ | (L) | (2) |
| $T_{ij}$ | Amount of Traffic between Country $i$ and $j$ | (G) | (2) |

IV. INFLUENCE BY THE COLD WAR

Since there was the Cold War during the pandemic, we consider the effect of the war. The Cold War was a period of political and economic antagonism between Western nations and Eastern nations. Influenza Pandemic of 1957-1958 occurred in the standoff period in the Cold War. There was no remarkable military confrontation. Thus we don't consider any military traffic for the international traffic.

However, the Cold War much influenced the international relationship between Western and Eastern nations, particularly in trade. Both Western and Eastern nations tended to have relationship with only same block countries. For example, according to [29][30], the Top 5 trading partners of the United States in 1957 were Canada, West Germany, United Kingdom, Japan, and Mexico, while those of USSR were East Germany, China, Czechoslovakia, Poland, and Romania. Thus many countries in the world had relationships with particular countries that belonged to same block. According to [29], it seems that trade relationship among Eastern nations was closer during the Cold War.

China was one of few countries which had exchanges with both Western and Eastern nations. However, according to

[30][31], China had no direct exchange with the United States during 1952-1971, which covers the pandemic period.

## V. SIMULATION AND RESULTS

We select China as the origin of the pandemic, and then examine some countries strongly related to China in terms of amount of trade referring to [31] and select USSR, Japan, United Kingdom, and West Germany. Next we examine the strongly related countries with these 4 countries and find 14 countries; Australia, Canada, Czechoslovakia, East Germany, France, India, Italy, Mexico, Netherlands, New Zealand, Poland, Romania, Sweden, and United States. Thus we simulate these 19 countries which are 2 neighborhoods from China.

For the global traffic, since we don't have exact data on the number of travelers during that period, we refer to the amount of trade between countries, instead. We expect that the people's traffic is larger when the amount of trade is larger between two countries, since the number of people who engage in the shipping is larger. Thus we assume that the situation of trade shows the international relationship at that that time. Then we consider main trading partners of a country and its amount of trade to understand the relationship between countries referring to [29][30][31]. For the local infection, we use the actual population and population density of that time in each country referring to [5][29][30][31].

As the preliminary simulation, we set the parameter values so the number of death cases in our simulation result closely matches the real data in three countries whose data is available as shown in Table 1; the United States, Japan, and United Kingdom. Figure 5 shows the simulation result on the number of death cases comparing with the real data. The number of death cases in the simulation result almost corresponds to those of the real data for three countries.

Then we extend the number of countries simulated to 19 countries using same parameter values. Figure 6 shows the simulation result on the number of death cases by country. We divided into two figures due to the different scales. Figure 6 (a) shows the most significant 4 countries and Figure 6 (b) shows other 15 countries. According to the simulation result, China, India, USSR, and the United States were the most impacted countries. Since we don't have real data on the number of death cases in each country to compare with our simulation result, we look at the total number of death cases in the world. In general, it is estimated that the total number of death cases in the world is about 2 million [1][2][3][4]. In our simulation result, the expected number of total death cases in the world is about 1.98 million. Thus our result is reasonable in terms of the number of death cases in the world.

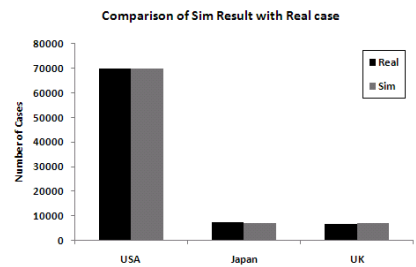

Figure 5: Comparison of the Number of Death Cases in 3 countries

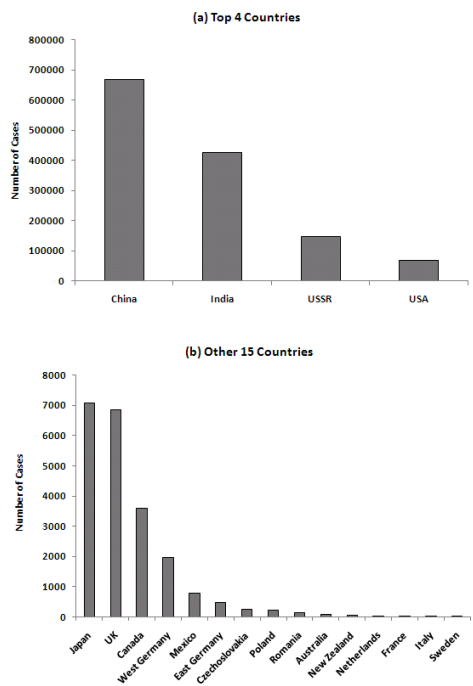

Figure 6: Number of Death Cases in 19 Countries in Simulation Result ((a) Top 4 Countries, (b) Other 15 Countries)

Figure 7 shows the infection routes that the simulation result shows. Note that the borders are current ones. Compared with the infection route in the real world shown as Figure 1, there are many corresponding points: Eastern Europe countries are infected through USSR. Western Europe countries are infected as another infection route. The United States are infected through Japan. Although, we don't consider Southeast Asian countries and Hong Kong in our simulation, when we consider these countries/regions, the infection route in the simulation may become closer to the real one.

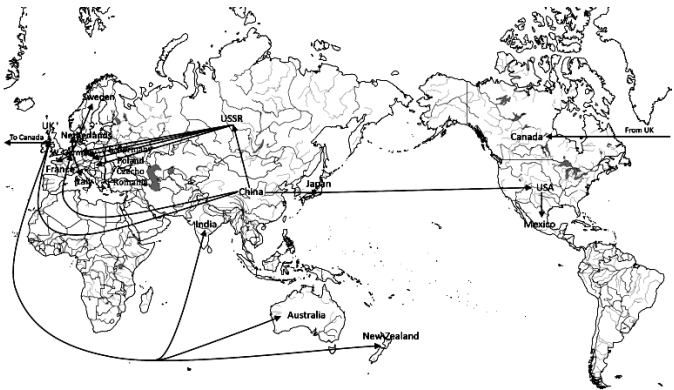

Figure 7: Infection Routes in Simulation

VI. SIMULATION WITH LESS EFFECT OF THE COLD WAR

We are interested in the influence of the Cold War upon the pandemic of 1957-1958. Thus we compare the result of the simulation for 1957-1958 with the result of another situation. Large transportation systems, including airplanes, had been developed after 1957; this is important because the development of transportation system has the potential to change not only the amount of trade, but also the speed of spreading of a pandemic. Thus we compare with periods before 1957. The world entered the Cold War period after the Second World War, and the effect lasted until the 1990's. During 1938 through 1945, the world was involved in the Second World War. During 1937 through 1945, China and Japan were engaged in the China-Japan War. We avoid these periods. Then we select two years; 1950 and 1936. In 1950, the Cold War had already begun. Also the Korean War broke out in June 1950. However, it is thought that the effect of the Cold War was still relatively insignificant in 1950 in terms of trade. The most remarkable difference between 1957 and 1950 is that there was still trade between the United States and China in 1950, while there were no such relations between these countries in 1957. In fact, the United States was the second largest trading partner of China in 1950 and the total amount of trade was 238 million US dollars, while the largest trading partner was USSR with the amount of 337 million US dollars [31]. According to [31], the trade between the United States and China decreased dramatically in the next year, and the relationship broke up after 1952. Thus we select the year of 1950 as the situation just before the relationship of the United States and China ended due to the Cold War. In addition, we pick the year of 1936, one year before the China-Japan War, assuming that trade was not influenced by any war in that year.

In order to determine the influence of the Cold War on the pandemic, we simulate another scenario for the comparison: the effect of the Cold War was less in 1957 and the international relationship during the pandemic should be like that during 1950 and/or 1936. Thus we refer to the relationship of trade between countries in 1950 and 1936, instead of 1957. However, the total amount of trade among 19 counties simulated in 1957 is larger than those of these years. The development of transportation system could influence the amount of trade. The difference of the total extent of trade likely influenced the simulation result. Thus we increase each amount of trade between two countries in 1950 and 1936 such that the total amount of trade within selected 19 countries equals to those of 1957. In addition, for the comparison, we just use the data on population and population density in 1957 since these data varies in years. Thus the total amount of trade, population, and population density are same with those for the simulation for 1957, and the only difference with the simulation for 1957 is the relationship between countries (i.e. each amount of trade between countries).

Then we simulate with the situation of 1950 and 1936 using same parameter values used in the previous simulation. Figure 8 shows the ratio of the number of death cases with each situation in countries to that with the situation of 1957. Figure 8 (a) is for 1950 and (b) is for 1936. When the ratio for a country is more than 1.0, the number of deaths with the situation of 1957 is smaller than that with the situation of another year. Thus that country would have larger number of death cases, if the effect of the Cold War is less. When the ratio for a country is less than 1.0, the number of deaths with the situation of 1957 is larger than that with the situation of another year. Thus that country would have smaller number of death cases, if the effect of the Cold War is less. In both Figure 8 (a) and (b), we omit France since its ratio is too high compared with other countries. This is because the number of death cases in France is very small with the situation of 1957. In Figure 8 (b), we show (unified) Germany, instead of West and East Germany since it was not divided in 1936.

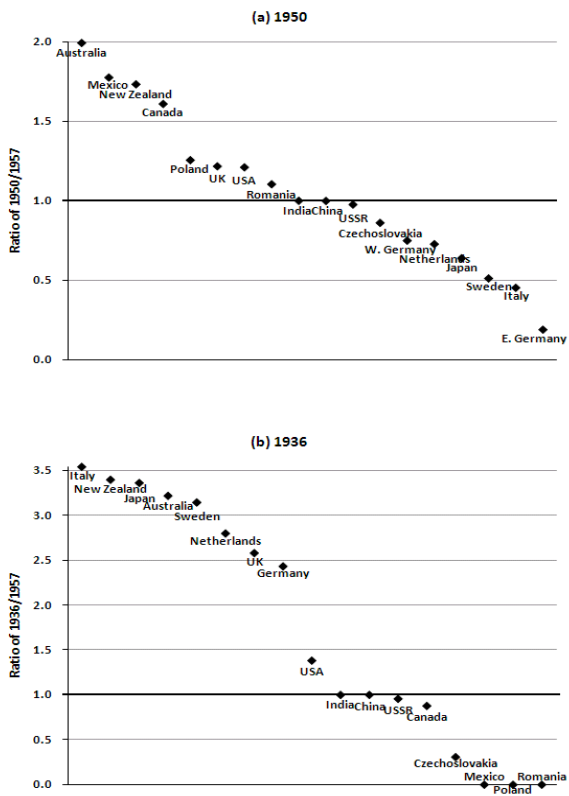

Figure 8: Ratio of the Number of Death Cases in 1950 and 1936 to that in 1957

From the results, a few observations may be made Western nations are more likely to have larger number of death cases, when the effect of the Cold War is less. Australia, France, New Zealand, United Kingdom, and the United States have ratio more than 1.0 with the situation of both 1950 and 1936. This is because the United States had exchange with China before 1950 and the pandemic spread quicker since the traffic is not restricted. In Figure 8 (a), Canada and Mexico have high ratio. Since the United States imports a case directly from China, these countries, which are adjacent to the United States, also infected earlier, which results in the larger number of local infections. Note that in Figure 8 (b), the ratio of these countries is not high since the amount of trade with the United States was not large in these countries in 1936.

On the other hand, Eastern nations, such as East Germany, Czechoslovakia, Poland, and Romania, are more likely to have smaller number of death cases, when the effect of the Cold War is less. This is more clearly shown in Figure 8 (b). Czechoslovakia and USSR have ratio less than 1.0 with the situation of both 1950 and 1936. This is because USSR had relationship with China and Eastern nations had the larger amount of trade with USSR during the Cold War. According to [29], the relationship among Eastern nations during the Cold War was closer than that of previous years in terms of the amount of trade. Thus it is believed that the Cold War strengthened the relationship among Eastern nations, especially with USSR, and that contributed to quick spread to Eastern nations through USSR and caused the larger number of number of death cases in Eastern nations.

For other counties, the ratio depends on the situation. In general, the increase/decrease of the number of death cases is influenced by the amount of trade with China or other countries which relate to China.

Next, we consider the impact on the world. Figure 9 shows the comparison of the expected total death cases in the world in different years. The impact in the world increases a little in different situations compared with that in the situation of 1957.

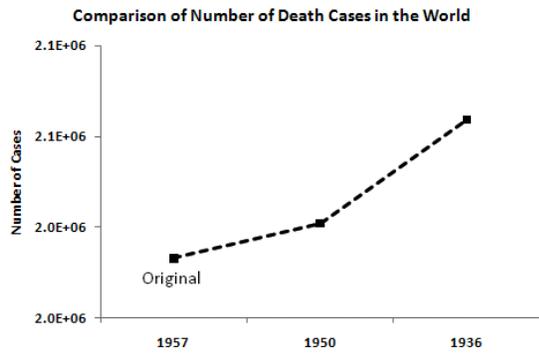

Figure 9: Comparison of Expected Total Deaths in the World

VII. CONCLUSIONS

In this paper, we simulated the Influenza pandemic of 1957-1958, considering the effect of the Cold War. International relationships were heavily influenced by the Cold War. We simulated, referring to the amount of trade, to understand the international relationship for the global infection and used the data on population and population density of the pandemic period for the local infection.

In order to explain the influence of the Cold War upon the pandemic, we compared the situation in 1957 with other years, 1950 and 1937. As the result, we found that if the effect of the Cold War was less, Western nations would have larger number of cases, and Eastern nations would have smaller number of cases. Thus the Cold War might work for the suppression of the pandemic for some countries, especially for Western nations, and the international relationship was an important factor for the spread of the pandemic.